\documentclass[aps,prd,showpacs,nofootinbib,twocolumn]{revtex4-1}
\usepackage{latexsym}
\usepackage{amsmath,amsfonts, MnSymbol}
\usepackage{hyperref}
\usepackage{amsbsy}
\usepackage{mathrsfs}
\usepackage{verbatim}
\usepackage{graphicx,calc,epsfig}
\def\ut#1{\rlap{\lower1ex\hbox{$\sim$}}#1{}}
\def\sdpb#1{\rlap{\lower1.5ex\hbox{$\Leftarrow$}}{#1}}

\newcommand{\C}{\mathbb{C}}
\newcommand{\R}{\mathbb{R}}

\newcommand{\be}{\nopagebreak[3]\begin{equation}}
\newcommand{\ee}{\end{equation}}
\newcommand{\bee}{\nopagebreak[3]\begin{equation*}}
\newcommand{\eee}{\end{equation*}}
\newcommand{\ba}{\nopagebreak[3]\begin{eqnarray}}
\newcommand{\ea}{\end{eqnarray}}
\newcommand{\baa}{\nopagebreak[3]\begin{eqnarray*}}
\newcommand{\eaa}{\end{eqnarray*}}
\newcommand{\la}{\label}
\DeclareFontFamily{U}{rsfs}{}         
\DeclareFontShape{U}{rsfs}{m}{n}{<5> rsfs5 <6><7> rsfs7          %
  <8><9><10><10.95><12><14.4><17.28><20.74><24.88> rsfs10}{}     %
\DeclareMathAlphabet{\mathfs}{U}{rsfs}{m}{n}                     %
\newcommand{\mfs}[1]{\mathfs {#1}}                               %
\newcommand{\va}{\scriptscriptstyle}

\newcommand{\n}{{\nonumber}}

\newcommand{\sH}{{\mfs H}}

\newcommand{\Hk}{{\sH}_{kin}}

\newcommand{\tr}{\mathrm{tr}}

\def\i{i}

\begin{document}

\title{Geometric temperature and entropy of quantum isolated horizons}


\author{Daniele Pranzetti$^1$}

\thanks{daniele.pranzetti@gravity.fau.de}

\affiliation{$^1$Max Planck Institute for Gravitational Physics (AEI), 
Am M\"uhlenberg 1, D-14476 Golm, Germany.}

\begin{abstract}
 
By reintroducing Lorentz invariance in canonical loop quantum gravity, we define a geometrical notion of temperature for quantum isolated horizons. This is done by demanding that the horizon state satisfying the boundary conditions be a  Kubo-Martin-Schwinger state. The exact formula for the temperature can be derived by imposing the reality conditions in the form of the linear simplicity constraints for an imaginary Barbero-Immirzi parameter.
Thus, our analysis reveals the connection between the analytic continuation to the Ashtekar self-dual variables and the thermality of the horizon.
 The horizon thermal equilibrium state can then be used to compute both the entanglement and the Boltzmann entropies. We show that the two provide the {\it same} finite answer, which allows us to recover the Bekenstein-Hawking formula in the semi-classical limit. In this way, we shed new light on the microscopic origin of black hole entropy by revealing the equivalence between  
the near-horizon degrees of freedom entanglement proposal and the state-counting interpretation. 
 
The connection with the Connes-Rovelli thermal time proposal for a general relativistic statistical mechanics is worked out.

\end{abstract}


\maketitle


\section{Introduction} \la{Introduction}

The analogy between black holes and thermodynamical systems is now well established. In particular, one can associate to a black hole a notion of entropy and temperature \cite{Bekenstein, Hawking}. However, the statistical mechanics understanding of these thermodynamical properties is not clear yet, as this would require most likely  a quantum gravity treatment of the horizon microscopic degrees of freedom (d.o.f.). Loop quantum gravity (LQG) has identified the entropy d.o.f. as the quantum fluctuations of the polymer-like geometry of the horizon \cite{SRK, ABCK, SU(2)} (see \cite{Review} for a review). While the result of the entropy counting showed a linear behavior with the horizon area (and the appearance of sub-leading logarithmic corrections) \cite{Counting},  the exact numerical match with the Bekenstein-Hawking formula $S = a_{\va H} /(4\ell^2_P)$ required to fix the Barbero-Immirzi parameter $\gamma$ (entering the spectrum of the area operator) to a particular real value.
The need to eliminate a purely quantum ambiguity represented by $\gamma$ through the request of agreement with a semiclassical result may be seen as a not very natural, let alone elegant, passage.

Indeed, this feature of the LQG calculation has received quite some attention during the years and some proposals for its removal have appeared (see, e.g., \cite{Jacobson-Ren, Dist}). However, only recently a key observation was made: in all of the state counting techniques developed in the literature the notion of temperature was never explicitly used. This observation led to the local stationary observer description of the horizon properties introduced in \cite{AP, APE}, which allowed to single out a physical notion of local horizon energy. Such an ingredient, together with the introduction by hand of the Unruh temperature \cite{Unruh} for the thermal atmosphere around the horizon, provides a leading term for the state counting independent on the Barbero-Immirzi parameter and in agreement with the Bekenstein-Hawking entropy \cite{AP}.

At the same time, the discovery of the important role played by the local description of the horizon physics led to the application of some techniques of state construction developed in the spin foam formalism \cite{SF} to the definition of a quantum Rindler horizon \cite{Bianchi}. In this context, the Unruh temperature is obtained by coupling a two-levels detector and then application of the Clausius relation yields the Bekenstein-Hawking formula for the entropy variation. 

It might seem quite surprising that the key role played by the horizon temperature in recovering the semiclassical result was realized only several years after the first ideas about black hole entropy in LQG appeared. However, it should be noted that the notion of Unruh temperature is intimately related to the local Lorentz invariance of space-time \cite{Bisognano, Haag}. On the other hand, in canonical LQG one uses the $SU(2)$ Ashtekar-Barbero connection \cite{Barbero} obtained from the Ashtekar's  complex variables \cite{Ash-con} by means of canonical transformations with real values of $\gamma$. The use of a compact group for the connection provides a rigorous definition of the kinematical Hilbert space $\Hk$ of the theory. 

However, differently from the $SL(2,\C)$ Ashtekar connection, which can be derived from a manifestly covariant action \cite{Jac-Smo} maintaining full local Lorentz invariance, the Ashtekar-Barbero connection cannot be interpreted as a space-time connection. In fact, it has been shown that it does not transform correctly under space-time diffeomorphisms \cite{Samuel, Alexandrov}. Only recently this issue has motivated the development of new techniques to reformulate the canonical theory in a Lorentz-covariant way keeping $\gamma$ real (see \cite{cov} for a review). These techniques provide new tools that will be used in our analysis, as explained below.  

In light of the strict connection between temperature and local Lorentz invariance, it seems necessary to start with a covariant approach to LQG in order to be able to {\it derive} a notion of temperature for the horizon. 
In this work we show that indeed, once Lorentz symmetry is introduced in the framework of quantum isolated horizons (QIH) \cite{ABCK, SU(2)},
a geometric notion of temperature arises. A possible way to implement Lorentz covariance is to introduce the natural
complex structure of $SL(2,\C)$ in the phase-space. That is, we reintroduce the complex Ashtekar variables, which, as argued above, are geometrically preferred (in the sense that they transform linearly under local Lorentz transformations). 
However, as recently shown in \cite{Wolfgang}, the use of an $SL(2,\C)$ connection does not necessarily imply an imaginary Barbero-Immirzi parameter. Therefore, 
we follow the approach of \cite{Wolfgang} and start from an $SL(2,\C)$ connection formulation while keeping $\gamma$ real. Then, the linear simplicity constraint of spin foam gravity
\be\la{simplicity}
K^i=\gamma L^i \,
\ee
emerges as a reality condition for the metric to be imposed in order to recover the standard $SU(2)$ kinematical
Hilbert space of LQG. Here $L^i$ and $K^i$ represent respectively the generators of the rotation and corresponding boost subgroups of the Lorentz group. 
Imposition of the relation \eqref{simplicity} in the quantum theory then brings us back to the real theory and, at the same time, tells us how to embed the $SU(2)$ subgroup in the $SL(2,\C)$ one. The arbitrariness of this embedding is then fixed by going to the time-gauge, which allows us to have the same $SU(2)$ structure everywhere (at each node).

Our strategy can thus be summarized as follows. In order to introduce the concept of temperature, we need to define a local notion of time evolution. This can be achieved by means of local Lorentz invariance. Therefore, we start classically from a complex phase-space parametrized by the Ashtekar complex variables. In the quantum theory this then leads us to the $SL(2, \C)$ analog of the holonomy-flux algebra. On this larger [than the standard $SU(2)$] kinematical Hilbert space, we then need to impose the reality condition \eqref{simplicity}. This can be done using techniques developed in the spin foam context. Since the subspace formed by the space of solutions to (the quantum analog of)  \eqref{simplicity} can be mapped to the space of cylindrical function of the $SU(2)$ Ashtekar-Barbero connection \cite{Wolfgang}, we recover the standard kinematical Hilbert space of loop quantum gravity. Therefore, the condition  \eqref{simplicity} plays a fundamental double role. On the one hand, it allows us to rely on the well-behaved  structures of the $SU(2)$ LQG kinematical Hilbert space for the construction of the horizon state; on the other hand, it provides a natural notion of time evolution encoded in the boost Hamiltonian $K^i$. In this way, all the necessary ingredients to define the KMS-condition defining the horizon temperature are at hand. 

Notice that up to this stage $\gamma$  is still kept real. One of the main results of the paper is to show that, in order for the KMS-condition to be satisfied by the horizon state (built as a solution to the IH boundary conditions) and derive a physical temperature, we need to set $\gamma=i$. Hence, the analytical continuation to be performed involves the $SU(2)$ embedding into $SL(2,\C)$ defined by the imposition of the simplicity constraint and, consequently, the spectra of the geometrical operators. In particular, while the spectrum of the area operator becomes imaginary, as a consequence of this analytic continuation, we show that the final result for the entropy is real. 

In this way, our analysis sheds light also on the calculation of \cite{Complex}, where  the passage from the compact gauge group $SU(2)$ to the non-compact group $SL(2,\C)$ is performed by means of such an analytic continuation of the Chern-Simons (CS) Hilbert space formula.

In Sec. \ref{sec:KMS}, in analogy with the Euclidean path integral representation of the Unruh vacuum,
we construct a KMS-state \cite{Haag} for the geometry of QIH associated to a proper sub-algebra of the holonomy-flux $^*$-algebra. The resulting local notion of temperature bears a geometrical characterization in terms of the conical singularities in the quantum horizon geometry induced by the punctures
 and shows the necessity of the condition $\gamma=i$. 

In Sec. \ref{sec:entropy} the KMS-state will be used to compute both the entanglement and the Boltzmann entropies of QIH, showing how the two provide the same finite answer which allows us to recover the Bekenstein-Hawking formula in the semi-classical limit. 
Interpretation of our results and comments will be presented in the final Sec. \ref{sec:final}.
In Appendix \ref{A} the relation with the Tomita-Takesaki modular theory \cite{Tomita} and the thermal time hypothesis of Connes and Rovelli \cite{Connes} is derived.


\section{ KMS-state and geometric temperature of QIH}\la{sec:KMS}

\subsection{General setup}\la{sec:set-up}

KMS-states correspond to the correct physical extension of Gibbs equilibrium thermal states to infinite dimensional quantum systems, thus providing the proper general formalism to define the notion of temperature.
In order to define a KMS-state associated to the geometry of QIH,
let us start by recalling the Euclidean path integral representation for the vacuum of a relativistic quantum field restricted to a Rindler wedge. It is in fact well known (see, for instance, \cite{thermal, thermal2}) that the thermality of a black hole \cite{GH} is analogue to the thermality of the vacuum in flat space-time when restricted to one wedge, namely the Unruh effect. More precisely, by means of an Euclidean path integral representation, the ground state wave functional of a field $\phi$ can be written as a reduced density matrix for the state restricted to the Rindler wedge. When expressed in terms of eigenstates $|n\rangle$ of the boost Hamiltonian $K_\eta$ generating the Rindler horizon in terms of shift of the hyperbolic angle $\eta$, the vacuum of a relativistic quantum field takes the form of the canonical thermal state
\be\la{Rindler}
\rho_{\va R}=\tr_L |0\rangle\langle0|=\sum_n e^{-2\pi E_n}|n\rangle_{\!\va R}{}_{\va R}\langle n|\,,
\ee
where $E_n$ are the eigenvalues of $K_\eta$ and $R, L$ refer to the modes on the right and left wedges. In this description one defines the horizon temperature as the period of the rotation angle corresponding to the euclidean continuation of $\eta$. Such a notion of temperature as the period of the Euclidean time coordinate in a set covering only part of the manifold can easily be extended to any static spacetime with a bifurcation horizon, and in particular to the Schwarzschild space-time. This is possible due to the fact that a wide class of metrics with horizons can be mapped to the Rindler form near the horizon; in the general case, the corresponding state is not a true vacuum, but it is invariant under the static Killing symmetry of the background.

Taking the trace of the reduced density matrix \eqref{Rindler} one obtains the partition function $Z$. In order to use this $Z$ to compute the entropy then, one has to go  off-shell \cite{Temp}. More precisely, the thermodynamic entropy is given by 
\be\la{ent}
S=-(\beta\frac{\partial}{\partial\beta}-1)\ln{Z}\,;
\ee  
therefore, in order to be able to derive with respect to $\beta$, $Z$ can be computed by evaluating the Euclidean functional integral over fields periodic in the Euclidean time under $t_E\rightarrow t_E+\beta$. This is achieved by extending the classical black hole action to conical geometries; namely, one introduces a conical singularity in the $(r,t)$ plane at the point where the horizon is located. One can show that the deficit angle of the cusp $\delta=2\pi-\beta$ and the horizon area are canonically conjugate. By means of the effective action of a quantum field on the background metric with a conical singularity at the entangling surface, the expression \eqref{ent}, when evaluated for standard black hole solutions (i.e. $\delta=0$), reproduces the familiar Bekenstein-Hawking entropy \cite{Temp}. 

The entanglement entropy interpretation of this black hole entropy derivation, originally proposed in \cite{Bombelli}, is clear in the geometric formulation of \cite{Replica} (see \cite{Solo} for a review of these ideas). 

The basic ingredient in this derivation is represented by Lorentz invariance. The use of the Ashtekar-Barbero connection in the previous LQG literature has obfuscated this fundamental aspect. Hence, we now want to reintroduce Lorentz invariance in the canonical framework and use it to define a notion of horizon temperature.

A possible way to do this is to follow the approach of \cite{Wolfgang}, where a program of canonical quantization with
respect to complex variables, following standard LQG techniques, has been initiated. As outlined in the introduction, this does not require an imaginary Barbero-Immirzi parameter and, hence, $\gamma$ can be kept real. Then, 
the simplicity constraints \eqref{simplicity} naturally appear as reality conditions on the momentum variable. Borrowing techniques developed in the spin foam context in order to embed the $SU(2)$ kinematical Hilbert space via the notion of projected spin
networks \cite{Livine}, the weak imposition of \eqref{simplicity} provides a limiting procedure in which one can recover the usual kinematical Hilbert space of the real valued Ashtekar-Barbero connection  \cite{Wolfgang}. 
In this way, we can rely on the standard $SU(2)$ kinematical structures, even though we started with a complex connection. At the same time, the fact that the (projected) $SU(2)$ spin network states satisfy the reality condition \eqref{simplicity} will play a crucial role in showing the thermality of the horizon state. Our analysis in the rest of this section will hence proceed as follows.

The quantum horizon state will be constructed from the invariant subspace of the tensor product between $SU(2)$ spin-$j$ representation spaces associated to the bulk links piercing the horizon, in analogy with the standard construction \cite{SU(2)}. On each  $SU(2)$ representation space we then need to impose the IH boundary condition \cite{IH, SU(2)}
\be\la{boundary}
{F}^{i}_{ab}(A^{\va }) = -\frac{\pi (1-\gamma^2)}{a_{\va H}}\, {L}_{ab}^{i}\,,
\ee
where $F^i(A)$ is the curvature of the connection $A$, $a_{\va H}$ the IH cross-section area\footnote{Recall that an IH  is null hypersurface topologically $S^2 \times R$, i.e. foliated by a (preferred)
family of 2-spheres $H$ (at the intersection between the IH and the bulk space-like surfaces) and equipped with an equivalence class $[\ell^a]$ of transversal
future pointing null vector fields generating the horizon.} and ${L}^{i}$ an $SU(2)$ valued two-form constructed from the triad field. Since each spin-$j$ vector space itself can be written as a tensor product between a part (of the link) living in the bulk and one on the boundary (see below for more details), the horizon density matrix will be obtained by tracing over the boundary degrees of freedom. 

At this point, the last missing ingredient  in order to define and impose the KMS condition for the horizon state is the introduction of a notion of time evolution. This is where the original Lorentz group structure comes in hand. In fact, locally (at each link) it is always possible, by a proper gauge fixing, to align the internal directions to the space-time directions; in this way, we can locally identify the gauge symmetry (internal $SO(1,3)$ rotations) with the spacetime diffeomorphisms and hence introduce a local notion of time.
Such gauge fixing is provided by the time gauge, which allows us to write the $SL(2,\C)$ algebra in terms of the generators $L^i$ of the $SU(2)$ rotations and the corresponding boost generators $K^i$ entering in \eqref{simplicity}. 
In this way, we can identify the internal space with local Minkowski space-time and single out an internal time-direction, aligned to the space-like hypersurface normal.  

With this identification, the boost Hamiltonian $K^i$ can be interpreted as a time evolution generator, like in the case of a Rindler horizon, and used to define the algebra automorphism entering the definition of KMS condition. This is possible since the IH boundary conditions are specified by introducing a choice of null tetrads system which is compatible with and preserves the space-time foliation; moreover, a further gauge-fixing (adapted to the horizon) is possible such that the rotation generators $L^i$ on the horizon coincide with the flux operators on the r.h.s. of \eqref{boundary}, in which the space-time indices are pulled back to the boundary 2-sphere, i.e. the $(\theta,\phi)$ plane, (hence the same notation) and the boost generators $K^i$ live in the $(r,t)$ plane (via the identification with the (0,1) internal directions). Namely, this corresponds to the gauge where the tetrad $(e^I)$ is such that $e^0$ is aligned to the space-like surface normal (time gauge), $e^1$ is normal  to $H$, $e^2$ and $e^3$ are tangent to $H$ and the null tetrad reads $\ell^a=2^{-1/2} (e^a_0+e^a_1)$, $n^a=2^{-1/2} (e^a_0-e^a_1)$, and $m^a=2^{-1/2} (e^a_2+i e^a_3);$ the only non-trivial component of the r.h.s. of \eqref{boundary} is then $L^1_{ab}\propto e^2_a\wedge e^3_b$.

 Notice that this local notion of time evolution is enough for our purposes to construct KMS states, since the holonomy-flux subalgebra considered in the definition of the $C^*-$algebra in Sec.  \ref{algebra} below is formed by local operators defined at each puncture.


 Let us conclude with a couple of comments. First of all, the linear simplicity constraints will always be used inside expectation values in the following. Thus, only their weak imposition will be necessary, consistently with the `new' spin foam models. Secondly, since the bulk dynamics is frozen for a QIH (the lapse function vanishes on the horizon \cite{IH}), what we are interested in here is just the restriction to the $SU(2)$ kinematical Hilbert space and not a given spin foam vertex amplitude. In particular, we do not need to choose a specific map implementing \eqref{simplicity} weakly on the labelings of the unitary representations of $SL(2,\C)$ at each puncture. In this way, one may consider a bigger space of solutions than the one defined by the EPRL model \cite{flip}, like for instance in \cite{BO}.


\subsection{$C^*-$algebra}\la{algebra}

The starting point for the construction of a KMS-state is a $C^*-$algebra. As pointed out above, the imposition of the linear simplicity constraints \eqref{simplicity} allows us to use the real $SU(2)$ kinematical Hilbert space. Therefore, we are going to define our $C^*-$algebra as a sub-algebra $\mathcal A^{\va H}_{\va \Gamma}$ of the holonomy-flux $^*-$algebra of LQG \cite{AL} localized on the horizon $H$. For the construction of $\mathcal A^{\va H}_{\va \Gamma}$ we follow the analysis of \cite{Sahlmann}, where the authors study a modification of the Ashtekar-Lewandowski measure on the space of generalized connections and look for a representation of this algebra containing states that solve the quantum analog the boundary conditions \eqref{boundary}. In this way, the horizon degrees of freedom are now represented simply by elements of the flux-holonomy algebra of LQG. While the construction of \cite{Sahlmann} is not fully worked out in the SU(2) case for all relevant observables, for the  definition of $\mathcal A^{\va H}_{\va \Gamma}$ we only use a subset of observables, which are not affected by quantization ambiguities.

Let us consider a given boundary graph $\Gamma$, that is a set of $N$ edges piercing the horizon. As shown in \cite{Sahlmann}, holonomies on all contractible loops over the horizon are fixed and there are no local gauge invariant d.o.f. associated to them, in accordance with the expected topological nature of the boundary theory. Therefore, we do not consider them in the algebra $\mathcal A^{\va H}_{\va \Gamma}$. Holonomies on loops around punctures in $\Gamma$ are fixed by the modified measure on $H$ implementing the geometric condition \eqref{boundary}.

On the other hand, in the case $H$ being a topological sphere, holonomies that run between punctures (and the conjugate fluxes) represent the only d.o.f. lying on the horizon 
and independent states on the horizon are labelled by a single intertwiner between them \cite{Sahlmann}. In this way one recovers the $SU(2)$ intertwiner model of \cite{SU(2)}. These holonomies can be seen as an extension of the edges in $\Gamma$; they form an intertwiner representing d.o.f. not accessible to the exterior observer and we are thus going to trace over them.
In the next section we will see how the information encoded in the intertwiner structure has an imprint in the thermal correlations of the resulting density matrix.

The horizon fluxes are defined on a discrete set of surfaces formed only by those branes $\{p_{\va 1}\dots p_{\va N}\}$ which intersect one of the edges in $\Gamma$ \footnote{This restriction guarantees that the density matrix defined in \eqref{KMS} can be represented as a cyclic and separating vector on a Hilbert space (see Appendix \ref{A}).}. On each of these branes then the fluxes are quantized as
\be
\hat L_p=\oiint_p \hat L \,.
\ee
 The last fundamental ingredient comes from the TQFT structure of the solutions to the condition \eqref{boundary} found in \cite{Sahlmann}. Namely, assuming the convergence between Chern-Simons and LQG isolated horizons quantization (as suggested also by the results of \cite{2+1}), we take the level $k$ (an integer entering the Chern-Simons description of the boundary theory---see also footnote 3) as a cut-off for the $SU(2)$ irreps labeling the punctures Hilbert space. In this way the boundary operators are bounded. This completes the definition of the $C^*-$algebra $\mathcal A^{\va H}_{\va \Gamma}$.

\subsection{KMS-condition}\la{sec:KMS-condition}

Following the Euclidean path integral approach, we can write the QIH state as a pure state by considering for each puncture the tensor product between the associated inside and outside (of the horizon) states. Here by `inside state' we refer to the extension of the bulk link that lies along the horizon and we use the suffix $I$ for this part of the state. Due to the entanglement between these two components then, one can obtain a mixed state when restricting to the exterior. 
\begin{figure}[ht]
\centering
\includegraphics[scale=0.3]{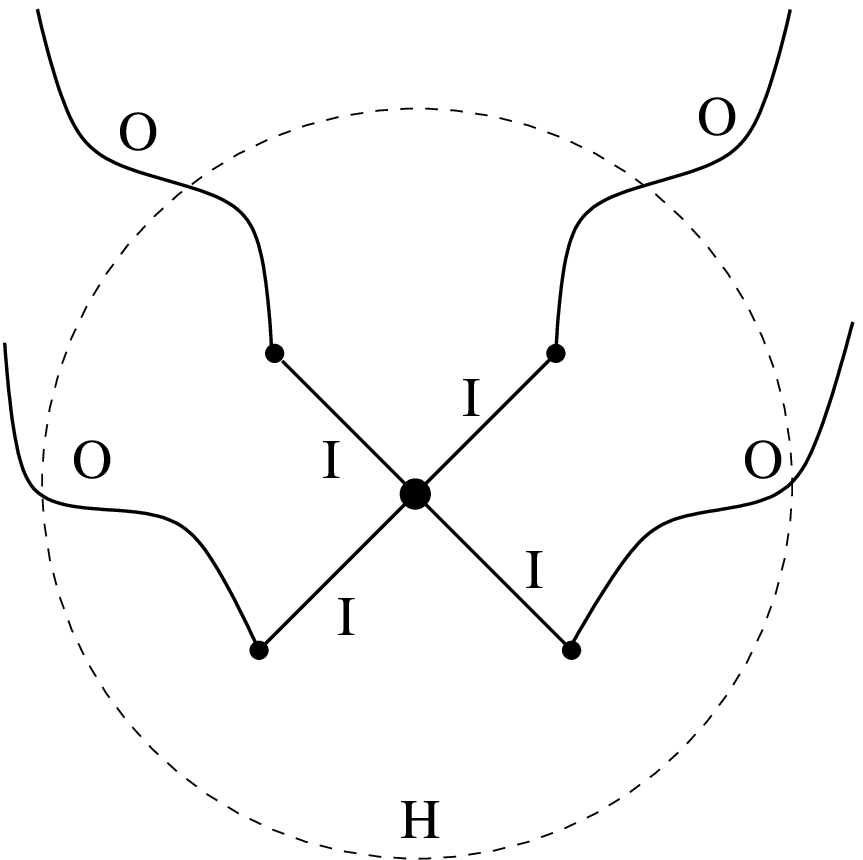}
\caption{Gauge invariant QIH state.}
\label{Punctures}
\end{figure}

Explicitly, by using the spin network basis spanning the kinematical Hilbert space of LQG, each link $p$ coming from the bulk, piercing the horizon and extending on its surface, like in FIG. \ref{Punctures}, has an associated irreducible representation space labelled by an $SU(2)$ spin $j_p$ (with $j_p=1/2,1, 3/2,\dots, k/2$).  The IH Hilbert space will then be given by a subspace of the tensor product of all such representation spaces associated to all the links piercing the boundary, namely $\sH_{\va IH}\subset \bigotimes_p |j_p \rangle$.
Let us now divide each boundary link Hilbert space in its internal and external parts by inserting a bivalent intertwiner at the point where the edge pierces the horizon (see FIG. \ref{Punctures}) and write the state associated to each puncture in the spin representation as
\be
|j_p\rangle=\sum_{m_p=-j_p}^{+j_p}|j_p,m_p\rangle_{\va I} \otimes|j_p,m_p\rangle_{\va O}\,,
\ee
where, due to orthogonality of spin network states,
\be\la{ortho}
{}_{\va A}\langle j_p,m_p|j'_{p'},m'_{p'}\rangle_{\va A'}=\delta_{pp'}\delta_{AA'}\delta_{j j'}\delta_{m m'}~~~{\rm with}~A,A'=I,O \,.
\ee
Thus, the horizon ``vacuum'' state belongs to the tensor product $\sH_{\va I} \otimes\sH_{\va O}$ between  the boundary and the bulk Hilbert spaces (as in the picture emerging from the Chern-Simons quantization \cite{ABCK, SU(2)}), and it can be written as a pure density matrix,
\be\la{vacuum}
\Omega=\bigotimes_p |j_p\rangle \langle j_p|\,,
\ee
on which the quantum boundary conditions have to be imposed.
In order to impose the quantum version of \eqref{boundary} on the state \eqref{vacuum}, we need to observe that the kinematical Hilbert space of the theory is given by cylindrical functions which depend on the connection via its parallel transport along links of a given graph. In other words, implementation of diffeomorphism invariance requires an initial configuration space containing holonomies and not directly the connection. Hence, to impose \eqref{boundary} in the quantum theory, we need to rely on the non-Abelain Stokes theorem relating surface integrals of the curvature with the holonomy (a similar regularization is required also in the Chern-Simons quantization of the boundary theory \cite{ABCK}). This amounts to exponentiating eq. \eqref{boundary} and imposing the quantum version of
\be\la{qboundary}
h_{\partial p} =\mathcal{P} \exp{i \oiint_p F(A)}=\mathcal{P} \exp{\frac{2\pi i}{k}\oiint_p L}\,,
\ee 
where we have used the linear relation between the Chern-Simons level $k$ and the IH area \cite{ABCK, SU(2)}\footnote{At this stage, the introduction of Chern-Simons ingredients like the level $k$ is not strictly necessary and one could as well replace $k$ with its expression in terms of the IH area $a_{\va H}$ and $\gamma$ following from \eqref{boundary}. However, in the following it will be important to use the cut-off $k/2$ on the spin representations labels in order to have finite traces over the states in the boundary theory.}. 

Imposition of \eqref{qboundary} in the quantum theory (which here we assume to be done like in \cite{Sahlmann} by means of a modified measure on the horizon Hilbert space) implies, in particular, that the state \eqref{vacuum} has to be gauge invariant. This means that the action of an holonomy that goes around all punctures has to leave the state invariant (given the 2-sphere topology of the IH, a loop surrounding all the punctures is contractible). Since this can be written as a product of holonomies around each puncture, we have
\be
\Omega=\bigotimes_p \hat h_{\partial p} |j_p\rangle \langle j_p|\,.
\ee
If we now trace over the degrees of freedom $|j,m\rangle_{\va I}$ inside the horizon and use the (quantum) boundary eom \eqref{qboundary} (by computing the scalar product on the internal states $I$ by means of the modified measure defined in \cite{Sahlmann}), the quantum statistical state is given by the density matrix
\be\la{KMS}
\hat\rho=\frac{1}{Z}\bigotimes_{p=1}^N P^{\va {j_{\va p}}}_{\va O}\, e^{\i\frac{2\pi}{k} \hat L_p}\,P^{\va {j_{\va p}}}_{\va O}= \frac{1}{Z}\bigotimes_{p=1}^N P^{\va {j_{\va p}}}_{\va O}\, e^{\i(\frac{2\pi}{k}-2\pi) \hat L_p}\,P^{\va {j_{\va p}}}_{\va O}\,,
\ee
where the projector $P^{\va {j_{\va p}}}_{\va O}$ is given by
\be\la{proj}
P^{\va {j_{\va p}}}_{\va O}=\sum_{m_p=-j_p}^{+j_p}|j_p,m_p\rangle_{\va O} {}_{\va O}\langle j_p,m_p|\,,
\ee
and 
\be\la{Z}
Z= \tr{(\bigotimes_{p=1}^N P^{\va {j_{\va p}}}_{\va O}\, e^{\i(\frac{2\pi}{k}-2\pi) \hat L_p}\,P^{\va {j_{\va p}}}_{\va O})}
\ee
is a normalization factor such that $\tr{(\hat\rho)}=1$.
The last passage in \eqref{KMS} follows from the periodicity of the action of the exponential of the flux operator \cite{Sahlmann}.

We can thus see how the information contained in the boundary conditions \eqref{boundary} is encoded in a Boltzmann-like factor on each puncture associated to the flux-Hamiltonian, in analogy with the Unruh vacuum \eqref{Rindler}, which emerges as a consequence of the imposition of diffeo-invariance and the trace over the inside states. In other words, the `Boltzmann' operator is not introduced by hand in the horizon state, but its appearance is a consequence of the QIH definition. 
Then, as it will be clearer in a moment, the thermality of the density matrix associated to the horizon quantum state originates from the entanglement between internal and external horizon d.o.f..

The trace-class operator $\hat\rho$ defines a positive linear functional (i.e. a state) over $\mathcal A^{\va H}_{\va \Gamma}$ that can be represented as
\be
\hat \rho (A)=\tr(A\hat\rho)
\ee
for every $A\in \mathcal A^{\va H}_{\va \Gamma}$.
Now remember that the state \eqref{Rindler} represents an example of KMS-state for the automorphism generated by the boost Hamiltonian over the algebra defined by functional calculi of the fields $\phi$ with support on the Rindler wedge \cite{Bisognano}. 
Therefore, based on the Bisognano-Wichmann theorem\footnote{The analogue of the Bisognano-Wichmann theorem, satisfied by the Unruh vacuum \eqref{Rindler}, for the case of the $C^*-$algebra defined in Sec. \ref{algebra} is shown to hold for the density matrix \eqref{KMS} in Appendix A. This provides further evidence for the local interpretation of \eqref{KMS} as the QIH `vacuum' state.} \cite{Bisognano} and the proposal of \cite{Bianchi}, using the gauge-fixing described in Sec. \ref{sec:set-up}, we can take the boost operator $\hat K$ on each brane as the local horizon generator. Hence, we consider the one-parameter automorphism group $\alpha_t$ generated by the boost Hamiltonian on the $C^*$-algebra $ \mathcal A^{\va H}_{\va \Gamma} $ localized on the horizon defined by
\be\la{auto}
\alpha_t(A)=\bigotimes_{p=1}^N P e^{\i t\hat K_{p}}P ~A \bigotimes_{p'=1}^N P' e^{-\i t\hat K_{p'}}P',~~A\in \mathcal A^{\va H}_{\va \Gamma}, ~t\in\R\,,
\ee
where from now on we adopt the notation $P\equiv P^{\va {j_{\va p}}}_{\va O}, P'\equiv P^{\va {j_{\va p'}}}_{\va O}$.
Using the GNS construction to define a representation associated to the state \eqref{KMS}, $\alpha_t$ can be shown to be the {\it modular automorphism group} on  $ \mathcal A^{\va H}_{\va \Gamma} $ associated to the state \eqref{KMS} (see Appendix \ref{A}). In this way then our analysis can be embedded in the formalism defined in \cite{Rovelli-stat} for a proposal of general relativistic statistical mechanics. From this point of view, the geometrical interpretation of the thermal time \cite{Connes} flow generated by the statistical state \eqref{KMS} is encoded in the boundary condition \eqref{boundary}.

If we now define the complex correlation function $f_{AB}(z)$ as
\be
f_{AB}(z)=\hat \rho (\alpha_z(A) B)~~~{\rm with}~ z\in\C\,,
\ee
 then the QIH state \eqref{KMS} satisfies the KMS-condition \cite{Haag}
\be\la{cond}
f_{AB}(-\i\beta)=\hat \rho (\alpha_{-\i\beta}(A) B)=\hat \rho (B\alpha_{0}(A))=f_{BA}(0)
\ee
with temperature  
\be\la{eta}
\beta=2\pi(1-\frac{1}{k})\,,
\ee
once the relation \eqref{simplicity} is implemented weakly on each brane with $\gamma=\i$. The reduced density matrix \eqref{KMS} then represents the analog of the KMS-state \eqref{Rindler} for QIH.

\vspace {0.15cm}
{\it Proof}:
\baa
&&f_{AB}(-\i\beta)\\
&&= \tr{\big(\bigotimes_{p'}\! P' e^{\beta \hat K_{p'}}\!P' A \bigotimes_{p''}\! P'' e^{-\beta\hat K_{p''}}\!P''B 
\bigotimes_p\! \frac{P e^{\i2\pi(\frac{1}{k}-1) \hat L_p}\!P}{Z}\big)}\\
&& =\tr{\big(B \bigotimes_p\! P e^{\i2\pi(\frac{1}{k}-1) \hat L_p}\!P
e^{\beta \hat K_{p}}\!P A \bigotimes_{p'}\! \frac{P' e^{-\beta\hat K_{p'}}\!P'}{Z}\big)}\\
&& =\tr{\big(B \bigotimes_p\! P e^{(\i2\pi(\frac{1}{k}-1)+\beta\gamma )\hat L_p}\!P
 A \bigotimes_{p'}\! \frac{P' e^{-\beta\hat K_{p'}}\!P'}{Z}\big)}\\
 &&= \tr{\big(B \bigotimes_p\! P A \bigotimes_{p'}P' \bigotimes_{p'}\! \frac{P' e^{\i2\pi(\frac{1}{k}-1) \hat L_{p'}}\!P'}{Z}\big)}\\
&&=\hat \rho (B\alpha_{0}(A)) = f_{BA}(0)\,,
\eaa
where the projectors' relation
\baa
&&\bigotimes_{p} P \bigotimes_{p'} P'\\
&&=\bigotimes_{p}\sum_{m_p}|j_p,m_p\rangle_{\va O} {}_{\va O}\langle j_p,m_p|
\bigotimes_{p'}\sum_{m_p'}|j_p',m_p'\rangle_{\va O} {}_{\va O}\langle j_p',m_p'|\\
&&=\bigotimes_{p}\sum_{m_p}|j_p,m_p\rangle_{\va O} {}_{\va O}\langle j_p,m_p|=\bigotimes_{p} P
\eaa
following from the orthogonality relation \eqref{ortho} has been applied several times;
in the third passage we have used \eqref{simplicity} and in the fourth one set
\bee
\gamma=\i~~~{\rm and}~~~\beta=2\pi(1-\frac{1}{k})\,.~\square
\eee

 The temperature \eqref{eta} has a geometrical interpretation in terms of the period of the rotational symmetry encoded in the action of the flux operator at each puncture. In fact, after imposition of \eqref{boundary}, the QIH state is flat except at the punctures $\{p\}_{\va 1}^{\va N}$ intersecting the surfaces over which $L$ is discretized, where it has conical singularities with quantized deficit angles $\delta=2\pi /k$ \cite{ABCK}.
We can thus see the further analogy with the Euclidean path integral approach. Namely, the notion of temperature emerging from the construction of KMS-states for QIH is associated to the quantum geometry deficit angle at the conical singularities in the boundary curvature induced by the bulk geometry. Since the r.h.s. of the boundary conditions \eqref{boundary} can be derived by means of the Gauss-Bonnet theorem \cite{ABCK}, such geometric connotation is analogue to the derivation of \cite{Temp}. More precisely, the notion of temperature derived from the IH boundary conditions is dual to the one of \cite{Temp} in the sense that, in our case,  the conical singularity lives in the $(\theta, \phi)$ plane instead of the $(r,t)$ plane (where we are using the correspondence between internal and space-time directions possible due to the gauge fixing described above). The equivalence between the two notions of temperature is then encoded in the relation between the (quantum) geometry of the two planes contained in \eqref{simplicity}. Hence, we see once more the important role played by the simplicity constraint. 

The relevance of the periodicity of the rotational symmetry around a puncture in the definition of the IH temperature is also supported by the conform field theory (CFT) description of the horizon d.o.f. provided in \cite{CFT}. In this case, the relation between the period on the circle (around a puncture) and that of the Euclidean time is encoded in the modular invariance of the CFT partition function on the torus. The origin of this match of the notion of temperature associated to the rotational symmetry on the horizon may be associated to the isomorphism between the (restricted) Lorentz group in four dimensions and the (projective) conformal group in two dimensions.

From the proof presented above it emerges clearly how the necessity of the condition $\gamma=\i$  is related to the kinematical structures used in LQG. More precisely, it is the passage from connection to holonomy variables in the quantum theory, which allows us to implement diffeomorphism invariance (as explained above), that brings in the factor $\i$ in the state \eqref{KMS} and ultimately requires  an imaginary $\gamma$ for the KMS-condition \eqref{cond} to be satisfied.
 Therefore, we can understand the physical meaning of the analytic continuation to the Ashtekar self-dual variables from a local point of view as a consequence of the deep relation between horizon temperature and local Lorentz invariance, as pointed out at the beginning.
 
Notice that the horizon boundary condition \eqref{boundary} was originally derived in \cite{IH} by means of the self-dual connection $A^{\va +}$, for which it takes the form $F(A^{\va +})=-2 \pi/a_{\va H}\, {L}^{\va + }$. Moreover, in the limit $\gamma=i$ the simplicity constraint \eqref{simplicity} becomes the self-dual condition $K=i L$, known to be part of the reality constraints of the Ashtekar complex formulation.
 
 The (dimensionless) Unruh temperature can be recovered by going to the large area semi-classical limit; in fact, in the large Chern-Simons level limit one gets immediately $\beta_{\va U}=\lim_{k\rightarrow\infty} \beta=2\pi$ \footnote{The usual formula $2\pi/\kappa$ for the Unruh temperature is recovered once we relate the flux-Hamiltonian to the horizon local energy of \cite{APE}, that is as we express the horizon evolution in terms of the static observer local time $t=\eta \ell$, where $\ell=\kappa^{-1}$ is the observer proper distance from the horizon. (Unfortunately, the standard notation might be confusing in this case: $k=$ Chern-Simons level, $\kappa=$ horizon surface gravity).}.

\section{Entropy}\la{sec:entropy}

We can now use the thermal equilibrium state $\hat \rho$ to compute the von Neumann and the Boltzmann entropies of QIH. 
Given the expression for the entanglement entropy of a reduced density matrix
\be
S_{ent}=-\tr (\hat \rho \ln{\hat \rho})\,,
\ee
from the state \eqref{KMS} 
and by means of the level $k$ as a natural IR cut-off on the punctures color to perform the trace (a UV cut-off is automatically implemented in the theory via the fundamental discrete structure of the quantum geometry), 
we get
\ba\la{ent}
S_{ent}&=&-\tr (\hat \rho \ln{\frac{1}{Z}\bigotimes_{p=1}^N P e^{-\i2\pi(1-\frac{1}{k}) \hat L_p}\,P)}\n\\
&=&\tr (\hat \rho \sum_p \i 2\pi(1-\frac{1}{k}) \hat L_p)  +\tr (\hat \rho \ln{Z})\n\\
&=&\i2\pi(1-\frac{1}{k})\,\tr (\hat \rho  \frac{ \hat a_{\va H}}{8\pi\i\ell_P^2})  +\ln{Z}\n\\
&=& \frac{\langle a_{\va H}\rangle}{4\ell_P^2 }(1-\frac{1}{k})+ N\ln{\tilde Z}\,,
\ea
where $Z=\tilde Z^N$
and in the third line we have used the analytic continuation to $\gamma=i$ of the formula for the LQG area operator, shown to be related to the gauge rotation generators by $A_p^2= (8\pi\ell_P^2)^2\gamma^2 L_p^2$ in \cite{flip}.

The Boltzmann entropy can be computed by noticing that the normalization \eqref{Z} of the horizon thermal state coincides with
the canonical partition function of QIH (see, for instance, \cite{AP}). By means of the result \eqref{eta} for the horizon temperature then we can use the formula
\be\la{Bol2}
S_{Bol}=-\beta^2\frac{\partial}{\partial\beta}\left(\frac{1}{\beta} \ln{Z}\right)\,.
\ee
Taking again $k$ as a regulator, we have
\ba\la{Bol}
S_{Bol}&=& -\beta\frac{\partial}{\partial\beta}\left( \ln{Z}\right)+\ln{Z}\n\\
&=&\beta\,\tr {(\hat \rho \sum_p \i \hat L_p) }+ N\ln{\tilde Z}\n\\
&=& \frac{\langle a_{\va H}\rangle}{4\ell_P^2 }(1-\frac{1}{k})+ N\ln{\tilde Z}\,,
\ea
where in the last passage we have used the expression \eqref{eta} for the temperature.

Hence, we see that the entanglement and the Boltzmann statistical entropies give the same result.
In the semi-classical limit of large horizon area (Chern-Simons level) we recover the Bekenstein-Hawking formula
\be
S_{BH}=S_{ent}=S_{Bol}
\ee
once the modification of the first law of QIH mechanics proposed in \cite{AP} is taken into account---such a modification encodes the quantum hair structure associated to the QIH geometry, as investigated more in detail in \cite{Evaporation2}. 

The appearance of this extra term can be better understood once taking into account also matter d.o.f. living on the horizon punctures. In fact, from the local observer perspective and the thermality of QIH derived here, it is natural to expect a distribution of matter fields quanta in a thermal bath near the horizon (see also the discussion in \cite{Evaporation2}). Such a ``gas'' will have its own entropy contribution to be added to the geometrical one \cite{Replica}. At large scales this quantum gravity local correction is expected to disappear due to the temperature redshift (and more in general the IR regime). A mechanism for such a disappearance has been proposed in \cite{Ghosh} using the renormalization argument of \cite{Jacobson-Ren}.  

Moreover, while the chemical potential contribution in  \eqref{ent} and \eqref{Bol} represents a quantum gravity correction, a similar term appears also in the statistical mechanical approach of \cite{Temp}, where the entropy is linked with the surface d.o.f. which give rise to `off-shell' conical singularities at the horizon for Euclidean black-hole geometries. As shown in \cite{Teit}, when determining the available phase space for the surface fields accounting for the entropy, the requirement of general covariance introduces a single dimensionless parameter $\mathcal N$ fixing the shape of the region of horizon phase-space available to the black hole. This $\mathcal N$, which enters the definition of the statistical ensemble, is thus associated to the volume of the horizon 2-sphere diffeomorphisms group and it yields a $\log{\mathcal N}$ correction to the Bekenstein-Hawking formula. 
 The possibility to associate a new thermodynamic parameter for black holes to this quantum mechanical correction was underlined already in \cite{Teit}.

To make sense of this quantum gravity correction in a purely gravitational context and be able to associate an observable to the quantum hair represented by the number of horizon punctures, a kinematical Hilbert space more akin to a Fock space than the usual LQG one seems to be required.  In this sense, the group field theory formalism \cite{Oriti} can provide the right framework to address this issue.

\section{Conclusions and comments}\la{sec:final}

We have seen that the reintroduction of Lorentz invariance in canonical LQG is a fundamental ingredient in order to define a notion of temperature for QIH.  
More precisely, we showed that the density matrix associated to the quantum horizon satisfies the KMS-conditon, i.e. it represents an equilibrium thermal state, with temperature given by \eqref{eta}. We saw that thermality requires the passage to an imaginary $\gamma$, revealing the connection between the Ashtekar self-dual variables and the thermality of the horizon and clarifying the proposal of \cite{Complex}.

Notice that the importance to restore Lorentz symmetry in the internal space relies on the fact that, by a proper gauge fixing, the space-time and internal symmetries can then be locally identified and this allows us to introduce a notion of local time evolution generated by the (internal) boost operator. This is a fundamental ingredient to define the temperature via the KMS-condition in Sec. \ref{sec:KMS-condition}. 
The special value $\gamma=i$ is not strictly necessary in order to introduce a complex structure in the phase-space; in fact, as stressed in the introduction,  we started from an $SL(2,\C)$ connection formulation with no constraint on $\gamma$. We then assumed that the reality condition \eqref{simplicity} has been imposed in order to recover (project into) the $SU(2)$ kinematical Hilbert space, which has been used to construct the horizon state \eqref{KMS}. 

However, due to the nice geometrical properties of Ashtekar self-dual connection, this can be regarded as the physically correct starting point of the theory. Hence, our point of view is to interpret the real $\gamma$ theory as an intermediate regularization procedure, since a well defined construction of the kinematical Hilbert space with an imaginary Barbero-Immirzi parameter from the beginning has so far been elusive. 
Then, $\gamma=i$ can be seen as the correct physical limit of the theory to be taken {\it at the end} of the calculation, and this expectation is supported, in our context, by the fact that such
specific value is in fact picked out by the requirement of the isolated horizon state to be a thermal state  (see also \cite{Bod} for further evidence).

 Therefore, this analytic continuation represents the analog of a procedure commonly applied in QFT, in which, in order for the calculations to be well defined, one first switches to Euclidean time and only at the end physical results are recovered by Wick rotating back to Lorentzian time. This Wick rotation interpretation is the sense in which the analytical continuation to $\gamma=i$ should be understood. A more rigorous way to perform this limit would consist of 
 realizing it as a phase-space Wick rotation, as analyzed in detail in \cite{Wick}\footnote{By means of the IH boundary conditions, it might be possible to find a simple expression, valid at the horizon, for the generator of the generalized coherent state transform defining the phase space Wick rotation between real and self-dual variables. In particular, an interesting case would be if this generator turn out be related to a boundary term in the action derived from the canonical analysis. Investigation in this direction is left for future work.}.

Our microscopic derivation reveals an intrinsically geometrical nature of the temperature related to the topological aspects of the boundary theory. 
Moreover, our proof of the KMS-condition also highlights how the horizon thermality originates from the imposition of the boundary condition \eqref{boundary} and the entanglement between internal and external d.o.f..

The expression of the QIH temperature \eqref{eta} presents a quantum correction $1/k$ which disappears in the semi-classical limit, when the Unruh formula is recovered, but that is expected to become relevant for microscopic black holes. 
However, since this correction encodes a sort of backreaction in the temperature formula, it might be relevant also for large black holes. More precisely, if we recall that the CS level takes the form \cite{ABCK, SU(2)} $k=a_{\va H}/a_P$, where $a_P$ is the Planck area, and use the local notion of horizon energy derived in \cite{APE}, then this quantum correction defines an effective temperature reproducing exactly the deviation from thermality of the radiation spectrum found in \cite{Wil}. Therefore, besides  showing consistency also with the semi-classical treatment of black hole radiance, our analysis may have important implications for the information paradox. Investigation of this quantum correction effect on the Hawking radiation spectrum along the lines of the analysis performed in \cite{Evaporation, Evaporation2} is left for future work.

Our identification of the QIH state as a KMS-state over a proper restriction of the LQG holonomy-flux $^*$-algebra to the horizon also shed new light on the microscopic nature of black hole entropy. In fact, as shown in Sec. \ref{sec:entropy}, the horizon KMS-state allows us to compute both the entanglement and the Boltzmann entropies. The results \eqref{ent} and \eqref{Bol} show how the two coincide. At the heart of this equivalence lies the derivation of the horizon temperature \eqref{eta} and the consequent fact that the partition function \eqref{Z}, obtained by tracing over the internal states d.o.f., coincides with the one used in the canonical ensemble calculation corresponding to the counting of the CS Hilbert space dimension \cite{AP}; this allows for  the identification of \eqref{Bol} as the Boltzmann entropy.

From a physical point of view, this correspondence can be understood as a relation between the quantum geometry d.o.f. correlations across the horizon encoded in $\hat \rho$ and the number of horizon `quantum shapes' encoded in the intertwiner space.  The duality between these two descriptions of the horizon quantum geometry
 originates from the fact that both the thermality of the reduced density matrix \eqref{KMS} and the intertwiner structure of the boundary Hilbert space are consequences of the imposition of the boundary condition \eqref{boundary}.

In this way, our analysis reveals an intriguing equivalence between  
the original near-horizon quantum fields entanglement proposal of \cite{Bombelli, Frolov} and the state-counting interpretation of black hole entropy \cite{SRK, ABCK}. This correspondence deserves further investigation to be understood more deeply and eventually be generalized to other approaches.

To conclude, let us point out how our work is based on an hybrid quantization procedure relying on elements of both the LQG and Chern-Simons formalisms. A description of QIH purely in terms of LQG structures, along the lines of the program initiated in \cite{Sahlmann}, would be desirable in order to put our results on a more solid ground. At the same time though, we have shown a nice interplay between techniques developed both in the canonical and covariant formalisms of the theory. In this way, black hole entropy calculation provides a clear example of how insights gained in one formulation of LQG can be fruitfully applied to the dual one.

\vspace {0.3cm}
{\it Acknowledgements.} It is my pleasure to thank C. Fleischhack, A. Perez, H. Sahlmann and T. Thiemann for helpful conversations. I am grateful to D. Oriti and H. Sahlmann for comments on a draft version of this manuscript that improved the presentation.

\begin{appendix}

\section{MODULAR AUTOMORPHISM GROUP}\la{A}

The Tomita-Takesaki modular theory \cite{Tomita} (see \cite{Bratteli} for a detailed technical description and \cite{Bertozzini} for an application as a possible approach to quantum gravity) provides a beautiful mathematical characterization of equilibrium states in statistical mechanics. Here we give a brief introduction and show its relation with the automorphism \eqref{auto} on the algebra $ \mathcal A^{\va H}_{\va \Gamma} $.

First of all, let us recall that, given an abstract $C^*$-algebra $ \mathcal A$ and a state $\omega$ over $ \mathcal A$, the GNS construction provides us with a Hilbert space $\sH$ with a preferred cyclic vector $|\Psi_0\rangle$ and a representation $\pi_\omega$ of $\mathcal A$ as a concrete algebra of operators on $\sH$ such that
\be
\omega(A)=\langle \Psi_0|\pi_\omega(A)|\Psi_0\rangle\,.
\ee

Now, let us consider a von Neumann algebra $\mathcal R$ acting on a Hilbert space $\sH$ possessing a cyclic ($A|\Omega\rangle$ dense in $\sH$) and separating ($A|\Omega\rangle=0~\rightarrow~A=0$) vector $|\Omega\rangle$. Consider the conjugate linear operator $S$ defined by:
\be
S A|\Omega\rangle=A^*|\Omega\rangle~~~~~~\forall A\in \mathcal R\,.
\ee
One can show that $S$ admits a polar decomposition,
\be\la{polar}
S=J \Delta^{1/2}\,,
\ee
where $\Delta$ is a self-adjoint positive operator and $J$ an antiunitary operator.
The Tomita-Takesaki theorem \cite{Tomita} states that the map $\sigma_t:\mathcal R\rightarrow \mathcal R$ defined by
\be\la{modular}
\sigma_t(A)=\Delta^{\i t} A \Delta^{-\i t}~~~~~~ A\in \mathcal R
\ee
defines a one-parameter group of automorphisms of the algebra $\mathcal R$, called the ``group of modular automorphisms'' of the state $\omega$ on the algebra $\mathcal R$. Correspondingly, $J$ is called the modular conjugation and $\Delta$ the modular operator of $(\mathcal R, \Omega)$. Then
\baa
&&J \mathcal R J= \mathcal R'\\
&& \Delta^{\i t} \mathcal R \Delta^{-\i t}   = \mathcal R\,,
\eaa
where $\mathcal R'$ is the set of all bounded linear operators on $\sH$ which commute with all elements of $\mathcal R$. 
It follows that the state $\omega$ is invariant under $\sigma_t$, i.e. $\omega(\sigma_t(A))=\omega(A)$ for all $A\in \mathcal R$ and $t\in \R$.
Moreover, one can show that $\omega$ satisfies the KMS-condition \eqref{cond} with respect to the automorphism group $\sigma_t$ for the inverse temperature value $\beta=1$. Therefore, an equilibrium state with inverse temperature $\beta$ may be characterized as a faithful state over the observables algebra whose modular automorphism group $\sigma_\tau$ (where $\tau=t/\beta$) is the time translation group. This is the key observation at the base of the thermal time hypothesis \cite{Connes}. Notice that, since a von Neumann algebra is also a concrete $C^{\va *}$-algebra, the Tomita-Takesaki theorem applies also to an arbitrary faithful state $\omega$ over an abstract $C^{\va *}$-algebra $\mathcal A$. This is due to the fact that $\omega$ defines a representation of $\mathcal A$ in terms of bounded linear operators on a Hilbert space via the GNS construction, as recalled above.
 
Following \cite{Haag} and the idea of \cite{Connes}, we are now going to use the GNS construction to define a representation of $ \mathcal A^{\va H}_{\va \Gamma} $ associated to the state \eqref{KMS} in which the map $\alpha_t$ given by \eqref{auto} can be shown to be the  modular automorphism group on it.
That the state $\hat\rho$ is cyclic can be easily seen from the properties of the holonomy-flux algebra. In order to be separating as well, we need to restrict the discretization of horizon fluxes to a set of surfaces intersecting once any of the boundary punctures. This was already taken care of in the definition of $\mathcal A^{\va H}_{\va \Gamma}$ in Sec. \ref{algebra}. Notice that all the links in the state \eqref{KMS} end on these surfaces due to the trace over the interior d.o.f. necessary to derive the density matrix; in this way, possible ambiguities and difficulties related to the non-commutative $SU(2)$ structure in the algebra disappear.

In order to derive the modular operator $\Delta$ from the polar decomposition \eqref{polar}, we need first to construct a new representation in which $\hat\rho$ is given by a vector $|\Omega\rangle$ in the new Hilbert space. This can be done by considering the set of Hilbert-Schmidt density matrices on the boundary Hilbert space $\sH_H$, that is the set $\{\kappa: \tr{(\kappa^*\kappa)}<\infty, ~\kappa\in \mathcal B(\sH_H)$\}, where $\mathcal B(\sH_H)$ is the set of all bounded, linear operators acting in $\sH_H$. This set forms a Hilbert space $\mathcal K$ with respect to the scalar product
\be
\langle\kappa|\kappa'\rangle=\tr(\kappa^* \kappa)\,.
\ee
Since the thermal state $\hat\rho$ is positive, the operator
\be
\kappa_0=\hat\rho^{1/2}
\ee 
can be seen as a pure vector in $\mathcal K$, which we denote $|\kappa_0\rangle$ and represents our cyclic and separating vector $|\Omega\rangle$. We can now consider the following representation of $\mathcal A^{\va H}_{\va \Gamma} \subset \mathcal B(\sH_H)$ by operators acting on $\mathcal K$,
\be
\pi(A)|\kappa\rangle=|A\kappa\rangle:~~~\kappa\in \mathcal K,~~A\in \mathcal A^{\va H}_{\va \Gamma}\,.
\ee
Due to the intersection property introduced above, $|\kappa_0\rangle$ is a cyclic and separating vector for the representation $\pi$. Moreover, one has
\be
\langle\kappa_0|\pi(A)|\kappa_0\rangle=\hat\rho(A)\,,
\ee
that is $|\kappa_0\rangle$ plays the role of the vacuum state vector in $\pi$, corresponding to the density operator $\hat\rho$.
Furthermore, the state $|\kappa_0\rangle$ is time invariant in the sense that, by representing the time flow generated by the automorphism \eqref{auto} as
\be
U(t)|\kappa\rangle=|\bigotimes_{p=1}^N P  e^{\i t\hat K_p}P~ \kappa \bigotimes_{p'=1}^N P' e^{-\i t\hat K_{p'}} P'\rangle\,,
\ee
relation \eqref{simplicity} implies
\be
U(t)|\kappa_0\rangle=|\kappa_0\rangle\,.
\ee 
We are now ready to identify the modular conjugation $J$ and operator $\Delta$ in the polar decomposition \eqref{polar} such that
\be\la{SA}
SA|\kappa_0\rangle=A^*|\kappa_0\rangle=|A^*\kappa_0\rangle\,,
\ee
namely
\be
J|\kappa\rangle=|\kappa^*\rangle
\ee
and
\be
\Delta^{1/2}|\kappa\rangle=|\bigotimes_{p=1}^N P  e^{-\frac{\beta}{2}\hat K_p} P~\kappa \bigotimes_{p'=1}^N P' e^{\frac{\beta}{2}\hat K_{p'}}P'\rangle\,.
\ee

Let us show that with this choice of $J$ and $\Delta$, \eqref{SA} is satisfied,
\baa
SA|\kappa_0\rangle&=&J\Delta^{1/2}|A \frac{1}{\sqrt{Z}}\bigotimes_{p}^N P\, e^{-i\pi(1-\frac{1}{k}) \hat L_p}\,P\rangle\\
&=&J|\bigotimes_{p'}^N\! P^{\va'} e^{-\frac{\beta}{2}\hat K_{p'}}P^{\va'} A
\frac{\bigotimes_{p}^N\! P e^{-i\pi(1-\frac{1}{k}) \hat L_p} P}{\sqrt{Z}}
\!\!\bigotimes_{p''}^N\! P^{\va''} e^{\frac{\beta}{2}\hat K_{p''}}P^{\va''}\rangle\\
&=&J|\bigotimes_{p'}^N\! P^{\va'}\, e^{-\frac{\beta}{2}\hat K_{p'}}\,P^{\va'} A
\frac{1}{\sqrt{Z}}\bigotimes_{p}^N\! P\rangle\\
&=&|A^*\frac{1}{\sqrt{Z}}\bigotimes_{p}^N P\, e^{-i\pi(1-\frac{1}{k}) \hat L_p}\,P\rangle\\
&=&A^*|\kappa_0\rangle\,,
\eaa
where we have used the analytic continuation of the the simplicity constraint \eqref{simplicity} to $\gamma=\i$ and the value $\beta=2\pi(1-\frac{1}{k})$ for the temperature, as found in Sec. \ref{sec:KMS-condition}. Finally, a similar calculation shows that the time evolution automorphism \eqref{auto} is related to the modular group \eqref{modular} by
\be
\sigma_t=\alpha_{\beta t}\,.
\ee
\end{appendix}


\end{document}